\documentclass[preprint,showpacs,pra,superscriptaddress]{revtex4-1}
\usepackage[latin9]{inputenc}
\usepackage{amsmath}
\usepackage{amssymb}
\usepackage{graphicx}

\makeatletter

\providecommand{\tabularnewline}{\\}

 \@ifundefined{textcolor}{}
 {%
   \definecolor{BLACK}{gray}{0}
   \definecolor{WHITE}{gray}{1}
   \definecolor{RED}{rgb}{1,0,0}
   \definecolor{GREEN}{rgb}{0,1,0}
   \definecolor{BLUE}{rgb}{0,0,1}
   \definecolor{CYAN}{cmyk}{1,0,0,0}
   \definecolor{MAGENTA}{cmyk}{0,1,0,0}
   \definecolor{YELLOW}{cmyk}{0,0,1,0}
 }

\makeatother

\begin{document}

\title{Long-Distance Trust-Free Quantum Key Distribution}


\author{Nicol\'o Lo Piparo} 

\affiliation{School of Electronic and Electrical Engineering, University of Leeds,
Leeds, UK}
\author{Mohsen Razavi}

\affiliation{School of Electronic and Electrical Engineering, University of Leeds,
Leeds, UK}

%




\begin{abstract}
The feasibility of {\em trust-free} long-haul quantum key distribution (QKD) is addressed. We combine measurement-device-independent QKD (MDI-QKD), as an access technology, with a quantum repeater setup, at the core of future quantum communication networks. This will provide a quantum link none of whose intermediary nodes need to be trusted, or, in our terminology, a trust-free QKD link. As the main figure of merit, we calculate the secret key generation rate when a particular probabilistic quantum repeater protocol is in use. We assume the users are equipped with imperfect single photon sources,
which can possibly emit two single photons, or laser sources to implement
decoy-state techniques. We consider
apparatus imperfection, such as quantum efficiency and dark count
of photodetectors, path loss of the channel,
and writing and reading efficiencies of quantum memories. By optimizing different system parameters, we estimate the maximum distance over which users can share secret keys when a finite number of memories are employed in the repeater setup. 
\end{abstract}


\maketitle



%

\section{Introduction}
%
%
%
%
Future quantum communications networks will enable secure key exchange among remote users. They ideally rely on user friendly access protocols in conjunction with a reliable network of core nodes \cite{QInternet_Kimble, QAccess_Toshiba, Razavi_MulipleAccessQKD}. 
For economic reasons, they need to share infrastructure with existing and developing classical optical communication networks, such as passive optical networks (PONs) that enable fiber-to-the-home  services \cite{Shields.PRX.coexist, Townsend_QI_home_2011}. The first generation of quantum key distribution (QKD) networks are anticipated to rely on a {\em trusted} set of core nodes \cite{secoqc, Sasaki:TokyoQKD:2011}. This approach, although the only feasible one at the moment, may suffer from security breaches over the long run. In the future generations of quantum networks, this trust requirement can be removed by relying on entanglement in QKD protocols \cite{Ekert_91, Biham:ReverseEPR:1996}. This can be facilitated via using the recently proposed measurement-device-independent QKD (MDI-QKD) \cite{Lo:MIQKD:2012, MXF:MIQKD:2012, MDIQKD_finite_PhysRevA2012, Pan_expMDIQKD_PRL2013} at the access nodes of a PON \cite{Razavi_IWCIT12} and quantum repeaters at the backbone of the network, as we consider in this paper. The former enables easy access to the network via low-cost optical sources and encoders, whereas the latter may rely on high-end technologies for quantum memories and gates. Both systems, however, rely on entanglement swapping, which makes them naturally merge together. More importantly, in neither systems would we need to trust the intermediary nodes that perform Bell-state measurements (BSMs). In this paper, we study the feasibility of such a {\em trust-free} hybrid scheme by finding the relationship between the achievable secret key generation rate as a function of various system parameters. We remark that this setup does not provide full device-independence but it removes the trust requirement from the intermediary network nodes that perform measurement operations. Our work provides insights into the feasibility of such systems in the future.
 The system proposed in \cite{Azuma:All_optical_QR_2013} combines MDI-QKD with quantum repeaters by using time reversed all photonic quantum repeaters. However, \cite{Azuma:All_optical_QR_2013} requires single photon sources as well as large cluster states. Instead, our scheme relies on conventional quantum repeaters, where entangled quantum memories are used to store qubits which are teleported to large distances through entanglement swapping. Moreover, users can use imperfect single-photon sources or lasers.
MDI-QKD is an attractive candidate for the access part of quantum networks. First, it provides a means to secure key exchange without trusting measurement devices. This is a huge practical advantage considering the range of attacks on the measurement tools of QKD users \cite{Qi:TimeShift:2007, Makarov:Bright:09, Wiechers:AftergateAttack:2011, Weier:DeadtimeAttack:2011}. Moreover, at the users' ends, it only requires optical encoders
driven by weak laser pulses. That not only makes the required technology for the end users much simpler, but also it implies that the costly parts of the network, including detectors and quantum memories, are now shared between all networks users, and are maintained by service providers. One final advantage of MDI-QKD is its reliance on entanglement swapping, which makes its merging with quantum repeaters, also relying on the same technique, straightforward. This will help us develop quantum networks in several generations, where the compatibility of older, e.g. trusted-node, and newer, e.g., our trust-free, networks can be easily achieved.In

Quantum repeaters are the key ingredients to trust-free networks. They traditionally rely on quantum memories (QMs) to store entangled states. In order to avoid the exponential decay of rate with channel length, in quantum repeaters, entanglement is first distributed over shorter distances and stored in QMs. Once we learn about the establishment of this initial entanglement, we can perform BSMs to extend entanglement over longer distances \cite{Razavi.Lutkenhaus.09}. Considering the complexity of joint operations needed for BSMs, as well as possible purification thereafter, quantum repeaters are anticipated to be developed in several stages. The first generation of quantum repeaters may rely on probabilistic approaches to BSMs, which can be implemented using linear optics devices \cite{DLCZ_01, Razavi.Amirloo.10, ProbReps:RevModPhys.2011, RUS_arXiv}. These systems expect to cover moderately long distances up to around 1000~km without the need for purification. In order to go farther we need to develop efficient tools for purification and deterministic BSMs as was initially envisaged in \cite{Zoller_Qrepeater_98}. Such deterministic quantum repeaters will replace the probabilistic setups once their technology is sufficiently mature. Finally, the most advanced class of repeaters are the recently proposed no-memory ones \cite{Munro:NatPhot:2012, Azuma:All_optical_QR_2013, Liang:NoMemRep_PRL2014}, in which, by using extensive error correction, one can literally transfer quantum states from one point to another.

In this paper, we focus on the probabilistic setups for quantum repeaters, and, among all possible options, we use the protocol proposed in \cite{Sangouard:single-photon:2007}, which relies on single-photon sources (SPSs). In an earlier work \cite{LoPiparo:2013}, we compared the performance of this protocol, which we refer to as the SPS protocol, in the context of QKD, with several other alternatives, once imperfections in the SPSs are accounted for. We found that under realistic assumptions, this protocol is capable of providing the best (normalized) key rate versus distance behavior as compared to other protocols considered in \cite{LoPiparo:2013}. The particular setup that we are going to consider in this paper is then a phase-encoded MDI-QKD setup, whose reach and rate are improved by incorporating a repeater setup, as above, in between the two users. It is worth noting that the easiest way to improve rate-vs-distance behavior is to add two quantum memories in the MDI-QKD setup \cite{Brus:MDIQKD-QM_2013, Panayi_NJP2014, Nicolo_paper2}. This approach will almost double the distance one can exchange secret keys without trusting middle nodes, but it is not scalable the same way that quantum repeaters are. It, nevertheless, provides a practical route toward building scalable quantum-repeater-based links.





The paper is structured as follows. In Sec.~II, we describe the main ingredients of our setup including the phase
encoding MDI-QKD and the SPS-based quantum repeaters. In Sec.~III, we present our methodology for calculating
the secret key generation rate for our hybrid system, followed by numerical results
in Sec.~IV. We draw our conclusions in Sec.~V.  
\section{Setup description}

\begin{figure}
\begin{centering}
\includegraphics[width=8.6cm]{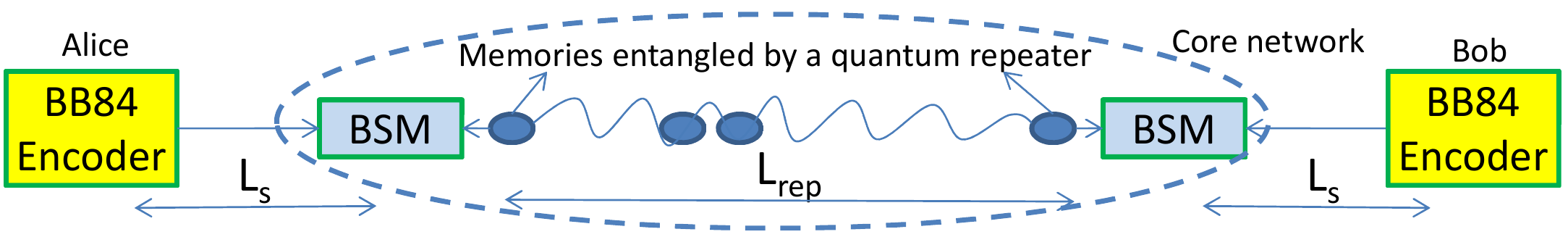}
\par\end{centering}

\caption{\label{fig:General-scheme}A general scheme for trust-free QKD links. Entangled states are created between internal nodes of the core network using quantum repeaters. The two BSMs will then enable an end-to-end MDI-QKD protocol.}

\end{figure}

In this section we first introduce the general idea behind our trust-free architecture and, then, explain particular MDI-QKD and quantum-repeater protocols considered for its implementation. Let us first consider the ideal scenario considered in Fig.~\ref{fig:General-scheme}. In this scheme, by using quantum repeaters, we distribute (polarization) entanglement between two memories apart by a distance $L_{\rm rep}$. This operation is part of the core network and is facilitated by the service provider. On the users' end, each user is equipped with a BB84 encoder, which sends polarization-encoded single photons to a BSM module at a short distance $L_s$ from its respective source. This resembles the access part of the network, where the BSM module is located at the nearest service point to the user. For each transmitted photon by the users, we need an entangled pair of memories to be read, i.e., their states need to be transferred into single photons. These photons will then interact with the users' photons at the two BSMs in Fig.~\ref{fig:General-scheme}.

The setup of Fig.~\ref{fig:General-scheme} effectively enables an enlarged MDI-QKD scheme. In MDI-QKD, the two photons sent by Alice and Bob are directly interacting at a BSM module \cite{Lo:MIQKD:2012}. Here, by the use of entangled memories, it is as if the Alice's photon is being {\em teleported} to the other side, and will interact with the Bob's photon at the second BSM. The overall effect is, nevertheless, the same, and once Alice and Bob consider the possible rotations in the memory states corresponding to the obtained BSM results, they can come up with correlated or anti-correlated bits for their sifted keys. Post processing is then performed to convert these sifted keys to secret keys.

\begin{figure}
\begin{centering}
\includegraphics[width=8.6cm]{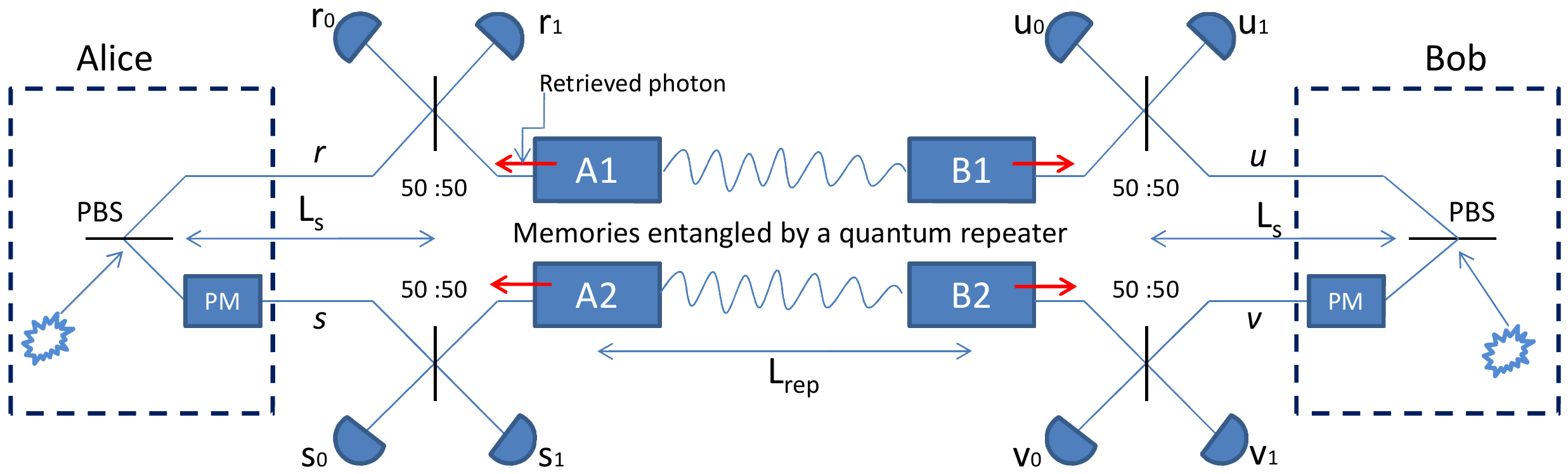}
\par\end{centering}

\caption{{\footnotesize \label{fig:Diagram-for-MDI-QKD}}Schematic diagram for a trust-free QKD link based on phase encoding. Memories are entangled using the SPS repeater protocol. Here, PBS stands for polarizing beam splitter and PM stands for phase modulator.}
\end{figure}

The same idea as in Fig.~\ref{fig:General-scheme} can be implemented via phase-encoding techniques as shown in Fig.~\ref{fig:Diagram-for-MDI-QKD}. Here, for simplicity, we have considered the dual-rail setup. The equivalent, and more practical, single-rail setup can also be achieved by time multiplexing as shown in \cite{MXF:MIQKD:2012}. In Fig.~\ref{fig:Diagram-for-MDI-QKD}, the quantum repeater ideally leaves memories $A_i$-$B_i$, for $i=1,2$, in the state $|\psi_{\rm ent}\rangle_{A_i B_i} = |0\rangle_{A_i}|1\rangle_{B_i} + |1\rangle_{A_i}|0\rangle_{B_i}$, where we have neglected normalization factors, and $|n\rangle_K$ represents $n$ excitations in memory $K$. The implicit assumption is that the memory is of ensemble type so that it can store multiple excitations \cite{Razavi.DLCZ.06}. The phase encoding that matches this type of entangled states is as follows. Alice and Bob encode their states either in the $z$ or in the
$x$ basis. Alice encodes her bits in the $z$ basis by sending, ideally, a photon in the $r$ or in the $s$ mode. This can be achieved by sending horizontally or vertically polarized pulses to the polarizing beam splitter (PBS) at the encoder. The same holds for Bob and his $u$ and $v$ modes. As for the $x$ basis, we can send a $+45^\circ$-polarized signal through the PBS to generate a superposition of $r\, (u)$ and $s\, (v)$ modes for Alice (Bob) state. Alice (Bob) encodes her (his) bits by choosing the phase value of the phase modulator (PM), $\phi_A \, (\phi_B)$, to be either 0 or $\pi$.

The BSMs used in the scheme of Fig.~\ref{fig:Diagram-for-MDI-QKD} are probabilistic ones. They will be successful if exactly two detectors, one from the top branch, and one from the bottom one, click. We recognize two types of detection. For the Alice's side (and, similarly, for the Bob's side), type I refers to getting a click on $r_0$-$s_0$ or on $r_1$-$s_1$. Type II refers to the case when $r_0$-$s_1$ or $r_1$-$s_0$ click. In order to get one bit of sifted key, Alice and Bob must use the same basis and both BSMs in Fig.~\ref{fig:Diagram-for-MDI-QKD} must be successful. Depending on the  results of these BSMs and the chosen basis by the two parties, Alice and Bob may end up with correlated or anti-correlated bits, where in the latter case, Bob will flip his bit. Table~\ref{Tab:bitassign} summarizes the bit  assignment procedure for our scheme. Note that these BSMs can be performed by untrusted parties.

\begin{table}
\begin{centering}
\vspace{2mm}
\begin{tabular}{|c|c|c|c|}

\hline 
Basis & Alice BSM & Bob BSM & Bit assignement\tabularnewline
\hline 
$z$ & type I/II & type I/II & Bob flips his bit\tabularnewline
\hline 
$x$ & type I (II) & type I (II) & Bob keeps his bit\tabularnewline
\hline 
$x$ & type I (II) & type II (I) & Bob flips his bit\tabularnewline
\hline 
\end{tabular}
\par\end{centering}

\caption{\footnotesize Bit assignment protocol depending on the results of the two BSMs. \label{Tab:bitassign} }
\end{table}

The repetition rate for our scheme is a function of several factors. In order to do a proper BSM, for each photon sent by the users, there must be {\em two} entangled pairs of memories ready to be read. In principle, the fastest that we can repeat our scheme is the minimum of the maximum source repetition rate, $R_S$, and half the entanglement generation rate of the quantum repeater, $R_{\rm rep}/2$. The latter is a function of the number of memories in use \cite{Razavi_SPIE}. We therefore consider two regimes of operation. If $R_S > R_{\rm rep}/2$, we then run our encoders at a rate equivalent to $R_{\rm rep}/2$ and will look at the achievable key rate per QM used. If $R_S < R_{\rm rep}/2$, i.e., when for every photon sent, there will be more than two entangled pairs ready, then we run our scheme at the rate $R_S$ and will look at the key rate per transmitted pulse as a figure of merit.  

In the following, we describe the quantum repeater protocol used in our scheme as well as different types of (imperfect) sources that users may use. Later, we look at the above achievable key rates once certain imperfections are considered in our setup.

\subsection{Source Imperfections}
\label{Sec:Source}
In our work, we consider two types of sources for the end users. The first type, which we will use as a point of reference for comparison purposes, is an imperfect SPS, with the following output state
\begin{equation}
\rho_{j}^{(\rm SPS)}=(1-p)\,|1\left\rangle _{jj}\right\langle 1|+p\,|2\left\rangle _{jj}\right\langle 2|,\quad j=A,\, B\label{eq:init_dens_fock_state}
\end{equation}
where $p$ is the probability to emit two, rather than one, photons. In practical regimes of operation, $p \ll 1$, hence, in our analysis, we neglect the simultaneous emission of two photons by both sources. The second type of source considered is a phase-randomized coherent source, which will be used in the decoy-state version of the protocol. In this case, Alice (Bob) will send $\mu = |\alpha|^2$ ($\nu = |\beta|^2$) photons on average for her (his) main signal states. Other values will be used for decoy pulses. Our analysis here only considers the case when there are infinitely many decoy states in use, although in practice we expect to achieve the same performance by using just a small number of decoy states \cite{MDIQKD_finite_PhysRevA2012}.

\subsection{SPS Repeater Protocol}

\begin{figure}
\begin{centering}
\includegraphics[width=7cm]{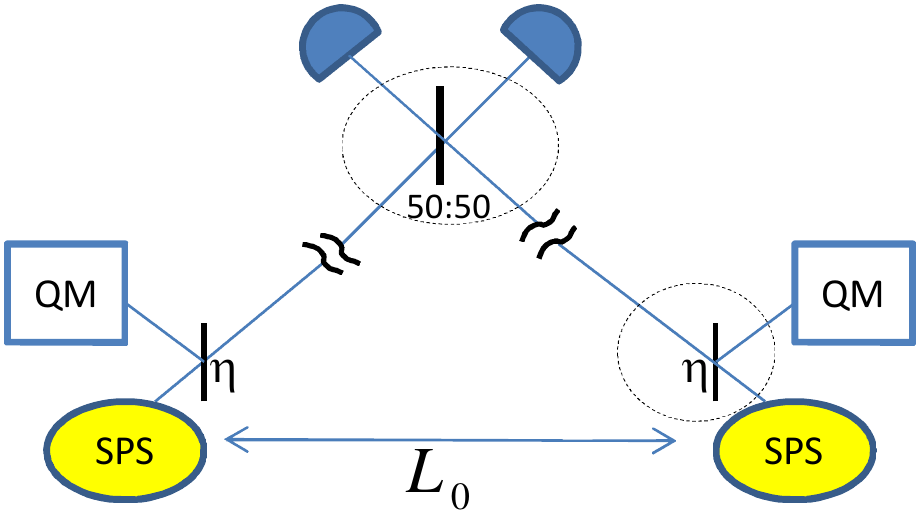}
\par\end{centering}
\caption{\label{fig:SPS-protocol}The SPS protocol for entanglement distribution.}
\end{figure}
The SPS protocol, proposed in \cite{Sangouard:single-photon:2007}, attempts to reduce the contribution of multi-photon
errors by using single-photon sources. The SPS setup for its initial entanglement distribution is shown in Fig.~\ref{fig:SPS-protocol}. In order to entangle two QMs at a distance $L_0$, corresponding to the shortest segment of the repeater setup, we send single photons through identical beam splitters with transmission coefficients $\eta$.
The photons can be reflected and stored in the QM or go
through the quantum channel and be coupled at a 50:50 beam splitter.
If exactly one of the two photodetectors in Fig.~\ref{fig:SPS-protocol} clicks, the memories are left in a mixture of an entangled state and a spurious vacuum term, where the latter can be selected out in later stages. For the entanglement swapping stage, we again use the 50:50 beam splitter followed by two single-photon detectors to perform a partial BSM.  
In \cite{LoPiparo:2013}, we calculate the secret key generation rate for the SPS protocol assuming that, instead of perfect SPSs, we are equipped with imperfect sources as in Eq.~\eqref{eq:init_dens_fock_state}. This is particularly a fundamental source of error, if one uses ensemble-based memories and the partial readout technique for generating single photons \cite{Sangouard:single-photon:2007}. Without loss of generality, we assume ensamble-based QMs with $\Lambda-$ level configuration and infinite decoherence time. The effect due to a finite decoherence time has been already considered in a previous paper \cite{Nicolo_paper2}. By considering writing and reading efficiencies for the QMs in use, respectively, denoted by $\eta_w$ and $\eta_r$, here we use the results of \cite{LoPiparo:2013} to find the relevant density matrices, $\rho_{A_i B_i}$ for $i=1,2$, for memories entangled by the SPS protocol for different values of $p$ and for different nesting levels $n$.
Other sources of imperfections considered throughout the paper are the path loss given by $\eta_{\rm ch}(l) = \exp(-l/L_{\rm att})$ with $L_{\rm att}$ being the attenuation length of the channel, photodetectors' quantum efficiency, $\eta_{d}$, and photodetectors' dark count per pulse given by $d_c$.

\begin{figure}
\begin{centering}
\includegraphics[width=8.6cm]{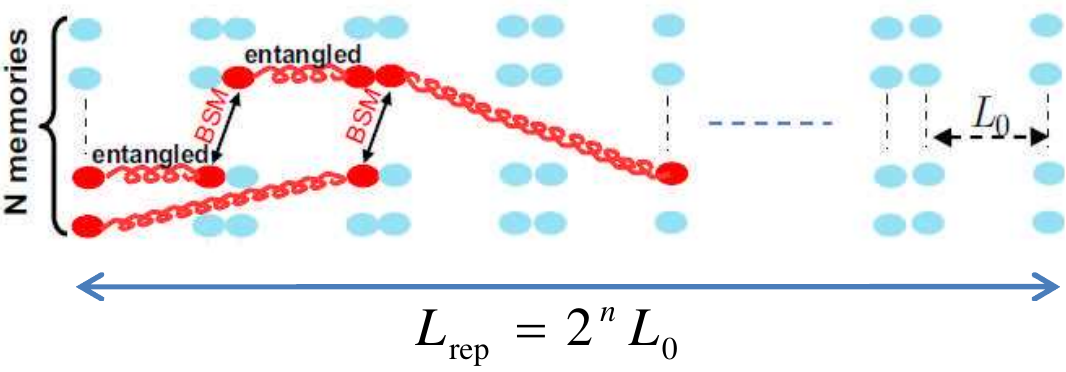}
\par\end{centering}
\caption{{\label{fig:(a)Multimemory-configuration-and}Multi-memory
configuration for quantum repeaters.} }
\end{figure}

In order to improve the entanglement generation rates in probabilistic quantum repeaters, it is essential to make use of multiple memories and/or multi-mode memories. Here, we assume a multi-memory structure as shown in Fig.~\ref{fig:(a)Multimemory-configuration-and} with $N$ memories per node, and employ the cyclic protocol proposed in \cite{Razavi.Lutkenhaus.09}. In this protocol, at each cycle of duration $L_{0}/c$ where $c$ the speed of light in the channel, we try to entangle, here using the SPS protocol, all the unentangled pairs of QMs at distance $L_{0}.$ At each cycle, we also perform as many BSMs as possible at the intermediate nodes. The main requirement for such a protocol is that, at the stations that we perform BSMs, we must be aware of establishment of entanglement over links of length $l/2$ before extending it to distance $l$ (informed BSMs). We use the results of \cite{Razavi.Lutkenhaus.09} to calculate the generation rate of entangled states {\em per memory} used, which is given by 
\begin{equation}
\begin{array}{c}
 R_{\mathrm{ent}}(L)=NP_{S}(L_{0})P_{M}^{(1)}P_{M}^{(2)}...P_{M}^{(n)}/T_{0}N2^{n+1}\\
=P_{S}(L_{0})P_{M}^{(1)}P_{M}^{(2)}...P_{M}^{(n)}/(2L/c)
\end{array}\label{eq:Rent}
\end{equation}
where $T_{0}$ is the duration of each cycle and $P_{S}\left(L/2^{n}\right)$ is the probability that the entanglement
distribution protocol succeeds over a distance $L_0$, $P_{M}^{(i)},i=1...n$,
is the BSM success probability at nesting level $i$ for a quantum
repeater with $n$ nesting levels. In our analysis, we
use the expressions for $P_{S}$ and $P_{M}^{(i)}$ up to two nesting
levels as found in \cite{LoPiparo:2013}. Finally, the total generation rate of entangled states in the limit of $N R_{\mathrm{ent}}(L) L/c \gg 1$ is given by
\begin{equation}
\label{eq:Rrep}
R_{\mathrm{rep}}(L)= N_{\rm QM} R_{\mathrm{ent}}(L),
\end{equation}
where $N_{\rm QM} = 2^{n+1}N$ is the total number of logical memories in Fig.~\ref{fig:(a)Multimemory-configuration-and}.

\section{secret key generation rate}

\begin{figure}
\begin{centering}
\includegraphics[width=7.5cm]{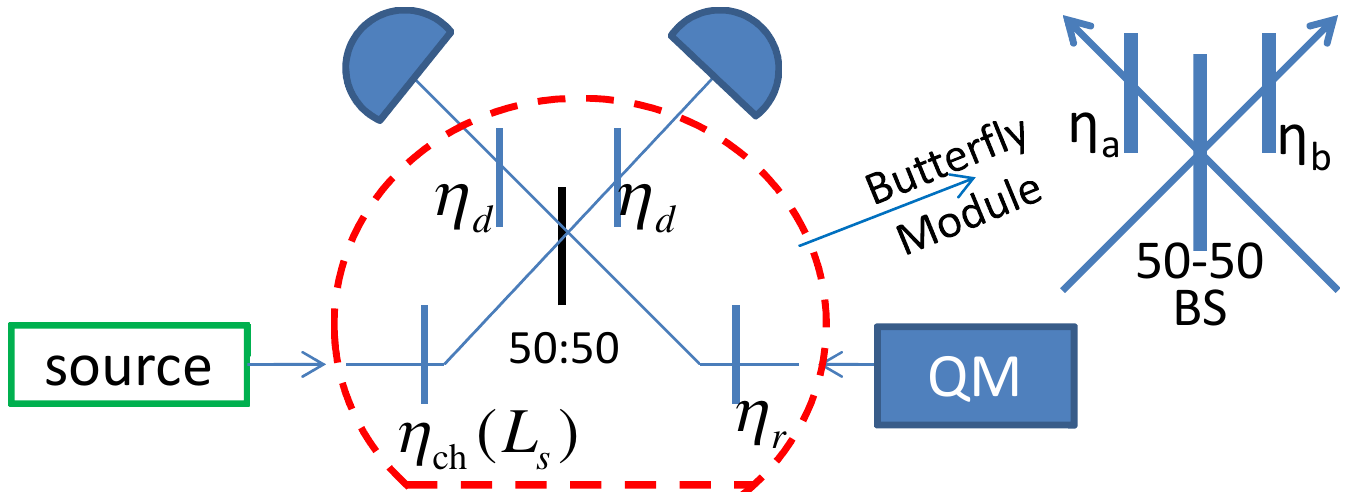}
\par\end{centering}
\caption{\label{fig:BSM-module}BSM module with generic transmission coefficient
represented by fictitious beam splitters. In our setup, $\eta_{a}$
is the path loss; $\eta_{b}$ is the reading efficiency and $\eta_{d}$
is the detection efficiency.}
\end{figure}

In this section, we find the secret key generation rate, $R_{\mathrm{QKD}}$,
per logical memory used, for the scheme of Fig.~\ref{fig:Diagram-for-MDI-QKD} under the normal mode of operation when no eavesdropper is present. We consider two types of sources as discussed in Sec.~\ref{Sec:Source}. 

\subsection{Imperfect SPSs}
Here, Alice and Bob each use an SPS with the output state as given by Eq.~\eqref{eq:init_dens_fock_state} in their encoder. In the limit of an infinitely long key and a sufficiently large number of QMs, their normalized secret key generation rate per employed memory is lower bounded by
\begin{equation}
\begin{array}{c}
R_{\rm QKD} = \frac{\min(R_S,R_{\rm rep}/2)}{N_{\rm QM}}\\
\times \mathrm{max}\left\{ Q_{11}^{z}\left(1-h\left(e_{11}^{x}\right)\right)-Q_{pp}^{z}f\, h\left(E_{pp}^{z}\right),0\right\} \label{eq:Rss}
\end{array}
\end{equation}
where $Q_{11}^{z} = (1-p)^{2}Y_{11}^{z}$, with $Y_{11}^{z}$ being the probability of a successful click pattern
in the $z$ basis when Alice and Bob send exactly one photon each; $e_{11}^{x}$ is the quantum bit error
rate (QBER) in the $x$ basis, provided that Alice and Bob are each sending exactly
a single photon; $Q_{pp}^{z}$ is the probability of a successful
click pattern in the $z$ basis when Alice and Bob use sources with outputs as in Eq.~\eqref{eq:init_dens_fock_state}, with the corresponding QBER given by $E_{pp}^{z}$; $f$ is the error correction inefficiency, and $h\left(x\right)=-x\,\log_{2}\left(x\right)-(1-x)\,\log_{2}\left(1-x\right)$
is the Shannon binary entropy function. 

Appendix \ref{AppA} provides us with the full derivation of the relevant terms in Eq.~\eqref{eq:Rss}. Our general approach to find these terms is as follows. For any basis $\Phi = x,z$ and any possible encoded state $\rho_{\rm enc}^\Phi = \rho_{rs} \otimes \rho_{uv}$ by Alice and Bob, the initial state of the system for memories $A_1$-$B_1$ and $A_2$-$B_2$ is given by
\begin{equation}
\rho_{\rm in}^\Phi = \rho_{\rm enc}^\Phi \otimes \rho_{A_1B_1} \otimes \rho_{A_2B_2} \label{eq:initial_dens_mat}
\end{equation}
where $\rho_{A_iB_i}$ has been obtained in \cite{LoPiparo:2013}. Once memories are read, their states will be transferred to photonic states, which we denote by the same label as their original memories. In that case, optical fields corresponding to modes $r$ and $A_1$, as well as the other three pairs of modes in Fig.~\ref{fig:Diagram-for-MDI-QKD}, would undergo through the setup shown in Fig.~\ref{fig:BSM-module}, where $\eta_a = \eta_r \eta_d$ and $\eta_b = \eta_{\rm ch}(L_s) \eta_d$. The equivalent sub-module in Fig.~\ref{fig:BSM-module} is what we refer to as an asymmetric butterfly module, whose operation is denoted by $B_{\eta_a \eta_b}^{ab}$ when it acts on two incoming modes $a$ and $b$. In \cite{Nicolo_paper2}, we have derived the output states of a butterfly module for relevant number states at its input. Using those results, we can then find the pre-measurement state right before the photodetection at the BSM modules by
\begin{equation}
\rho_{\rm out}^\Phi = B_{\eta_a \eta_b}^{rA_1} \otimes B_{\eta_a \eta_b}^{sA_2} \otimes B_{\eta_a \eta_b}^{uB_1}\otimes B_{\eta_a \eta_b}^{vB_2}  (\rho_{\rm in}^\Phi).
\end{equation}    
Note that we have already accounted for the quantum efficiency of photodetectors in our butterfly modules. The probability for a particular pattern of clicks on detectors $r_i$, $s_j$, $u_k$, and $v_l$, for $i,j,k,l = 0,1$, is given by
\begin{equation}
P_{r_{i}s_{j}u_{k}v_{l}}(\rho_{\rm enc}^\Phi)=\mathrm{tr}{\left(\mathit{\rho_{\rm out}^\Phi M_{r_{i}}M_{s_{j}}M_{u_{k}}M_{v_{l}}}\right),}\label{eq:Probability}
\end{equation}
where for $x=r,s,u,v$
\begin{equation}
\begin{array}{c}
M_{x_{0}}=(1-d_{c})\left[\left(I_{x_{0}}-|0\rangle_{x_{0}x_{0}}\langle0|\right)\otimes|0\rangle_{x_{1}x_{1}}\langle0|\right.\\
\left.+d_{c}|0\rangle_{x_{0}x_{0}}\langle0|\otimes|0\rangle_{x_{1}x_{1}}\langle0|\right]\label{eq:measurement_operators-1}
\end{array}
\end{equation}
is the measurement operator to get a click on detector $x_0$ but not on $x_1$. Here, $I_{x_0}$ denotes the identity operator for the mode entering detector $x_0$. One can define a similar operator $M_{x_1}$ by swapping subscripts 0 and 1 in the above equation. Hence, for example, looking at Fig.~\ref{fig:Diagram-for-MDI-QKD} the measurement operator corresponding to a click on detector $r_0$ and no click on $r_1$ is given by 
\begin{equation}
\begin{array}{c}
M_{r_{0}}=(1-d_{c})\left[\left(I_{r_{0}}-|0\rangle_{r_{0}r_{0}}\langle0|\right)\otimes|0\rangle_{r_{1}r_{1}}\langle0|\right.\\
\left.+d_{c}|0\rangle_{r_{0}r_{0}}\langle0|\otimes|0\rangle_{r_{1}r_{1}}\langle0|\right]
\end{array}
\end{equation}
The relevant terms in Eq.~\eqref{eq:Rss} can now be calculated by using Eq.~\eqref{eq:Probability} as shown in Appendix \ref{AppA}.

\subsection{Coherent sources}

In this section we replace the SPSs with lasers sources and use the decoy-state technique to exchange secret keys. This is a more user friendly approach as the complexity of the required equipment for the end users would be minimized. In the limit of infinitely many decoy states, infinitely long key, and sufficiently large number of memories, the secret key generation rate per logical memory used is lower bounded by
\begin{equation}
\begin{array}{c}
R_{\rm QKD} = \frac{\min(R_S,R_{\rm rep}/2)}{N_{\rm QM}}\\
\times \mathrm{max}\left\{ Q_{11}^{z}\left(1-H\left(e_{11}^{x}\right)\right)-Q_{\mu\nu}^{z}f\, H\left(E_{\mu\nu}^{z}\right),0\right\}, \label{eq:Rcc}
\end{array}
\end{equation}
where $Q_{\mu\nu}^{z}$ is the probability of a successful click
pattern in the $z$ basis when Alice and Bob send phase-randomized coherent pulses,
respectively, with mean photon number $\mu = |\alpha|^{2}$ and $\nu = |\beta|^{2}$ and
$E_{\mu\nu}^{z}$ is the QBER in the $z$ basis in the same
scenario. 

The procedure to find $Q_{\mu\nu}^z$ and $E_{\mu\nu}^z$ is the same as what we outlined in Eqs.~\eqref{eq:initial_dens_mat}-\eqref{eq:measurement_operators-1}. The only difference here is that in our butterfly modules, we now need to know the output of the module to coherent states in one input port, for the signal coming from the users, and number states in the other, representing the state of QMs. Table~\ref{tab:coherent} in Appendix~\ref{AppA} provides us with the input-output
relations for a range of relevant input states. We can then find the relevant terms of the key rate, as shown in Appendix~\ref{AppA}.

\section{Numerical Results}

In this section, we present numerical results for the secret key generation rate of our long-haul trust-free QKD link versus different system parameters. We look at two regimes of operation; the {\em source-limited} regime when memories are abundant and we are slowed down by source rates, i.e., $2 R_S < R_{\rm rep}$, versus the {\em repeater-limited} regime when the rate limitations come from the quantum repeater side, i.e., $2 R_S > R_{\rm rep}$. In the latter case, we should still satisfy the condition $N R_{\mathrm{ent}}(L) L/c \gg 1$ in order that Eqs.~\eqref{eq:Rent}-\eqref{eq:Rrep} remain valid.
We have used Maple 15 to analytically derive expressions for Eqs. \ref{eq:Rss} and \ref{eq:Rcc}.
Unless otherwise noted, we use the nominal values summarized in Table \ref{tab:Nominal-values-used}.

\begin{table}
\begin{centering}
\vspace{2mm}
\begin{tabular}{|c|c|}
\hline 
Memory writing efficiency, $\eta_{w}$  & 0.78\tabularnewline
\hline 
Quantum efficiency, $\eta_{d}$  & 0.93\tabularnewline
\hline 
Memory reading efficiency, $\eta_{r}$  & 0.87\tabularnewline
\hline 
Dark count per pulse, $d_{c}$  & $10^{-9}$\tabularnewline
\hline 
Attenuation length, $L_{\rm att}$  & 25 km\tabularnewline
\hline 
Speed of light in optical fiber, $c$  & $2 \times 10^{5}$ km/s\tabularnewline
\hline 
Double-photon probability, $p$ & $10^{-4}$\tabularnewline
\hline 
Access network length, $L_s$ & 5 km\tabularnewline
\hline 
Error correction inefficiency, $f$ & 1.16\tabularnewline
\hline 
\end{tabular}
\par\end{centering}

\caption{{\footnotesize \label{tab:Nominal-values-used}Nominal values used
in our numerical results.}}
\end{table}

\begin{figure}
\begin{centering}
\includegraphics[width=8.6cm]{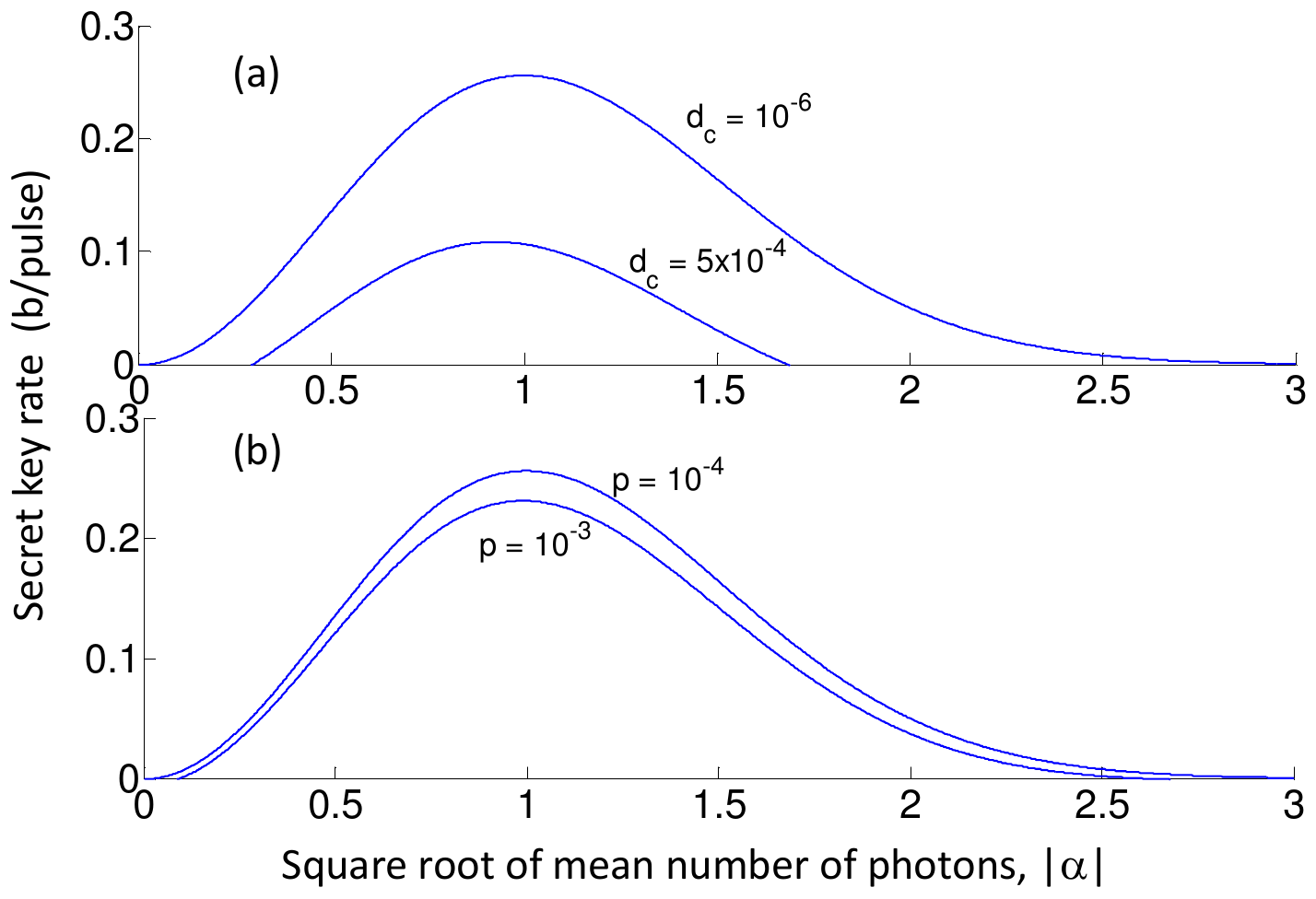}
\par\end{centering}

\caption{\label{fig:R_vs_alpha} Secret key generation rate per pulse versus 
$|\alpha| = |\beta|$ for different values of (a) the dark count and (b) the repeater's double
photon probability. Here, $L_{\rm rep}$ = 100 km and the other values are as in Tab. \ref{tab:Nominal-values-used}.}
\end{figure}

The first thing to obtain is the optimum intensity for our decoy-state scheme. Let us assume that in the symmetric scenario, as considered in this paper, Alice and Bob both use the same intensity value $\mu = |\alpha|^2 = \nu$ for their coherent signal states. Figure~\ref{fig:R_vs_alpha} shows the
secret key generation rate per pulse versus $|\alpha|$ 
for (a) different values of $d_{c}$ and (b) different values of $p$ of the quantum repeater
at $L_{\rm rep}=100$ km. We assume that $2 R_S < R_{\rm rep}$ and the plotted curves represent $R_{\rm QKD}N_{\rm QM}/R_S$ in Eq.~\eqref{eq:Rcc}. It can be seen in both figures that $|\alpha| = 1$ almost gives us the maximum rate in most scenarios. The optimal value is to some extent a function of $d_c$ as can be seen in Fig.~\ref{fig:R_vs_alpha}(a). By increasing $d_{c}$, the optimal intensity slightly
decreases. Dark count represents the main source of error in the $z$ basis, therefore, when
$d_{c}$ increases, the tolerance for the multiple-photon terms in a coherent state decreases, hence the maximum allowed value of $|\alpha|$ will go down as well. This leads to a
slightly shifted curve and therefore lower values for the optimal
values of $|\alpha|$. On the contrary, $E_{\mu\nu}^{z}$ is not
affected much by the double-photon probability $p$ and there is not much difference in the optimal intensity
when $p$ increases as shown in Fig. \ref{fig:R_vs_alpha}(b). We also obtain the same optimal values of $|\alpha|$ for  nesting levels one
and two in the repeater-limited regime. Throughout this section, we then use $|\mu| = |\nu| = 1$ in our calculations.
 
\subsection{Rate versus distance}

\begin{figure}
\begin{centering}
\includegraphics[width=8.6cm]{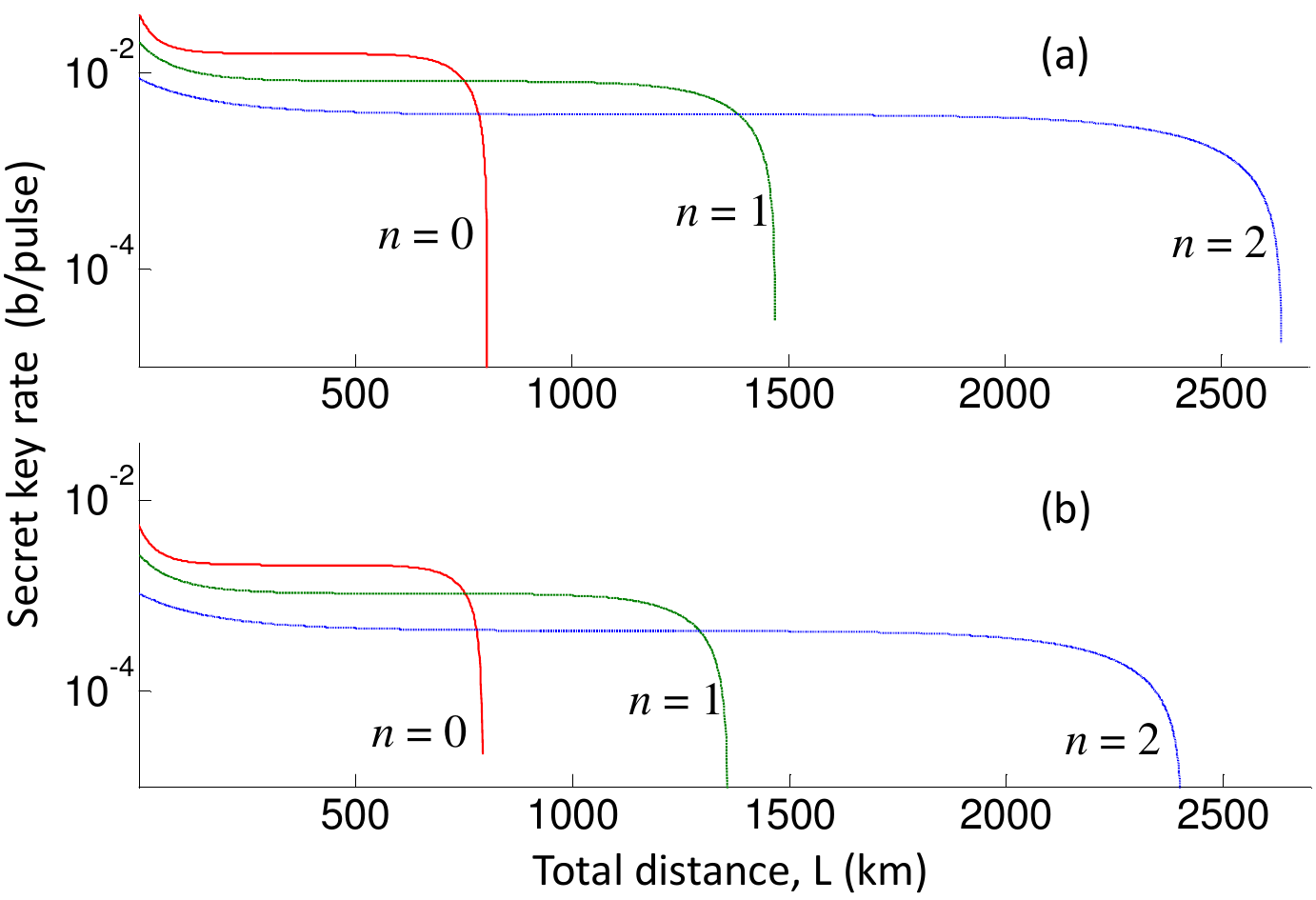}
\par\end{centering}
\caption{{\footnotesize \label{fig:R vs dist inf mem}Secret key generation rate per transmitted pulse, in the source-limited regime, versus
distance when (a) imperfect SPSs and (b) decoy coherent states
are used.}}
\end{figure}

\begin{figure}
\begin{centering}
\includegraphics[width=8.6cm]{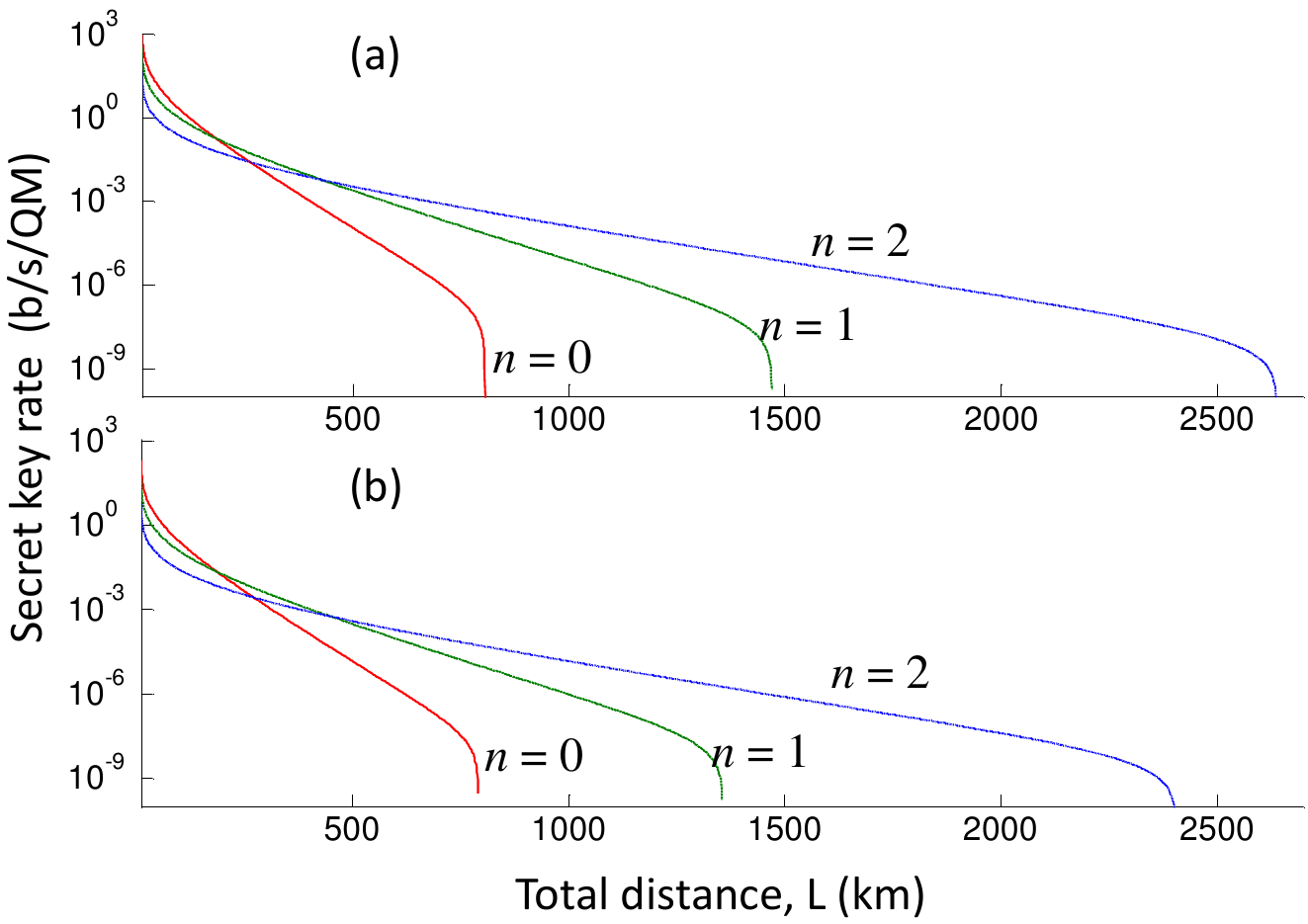}
\par\end{centering}
\caption{{\footnotesize \label{fig: R vs dist fin mem}$R_{\mathrm{QKD}}$, in the repeater-limited regime,
versus distance when (a) imperfect SPSs and (b) decoy coherent states are used. }}
\end{figure}

Figures \ref{fig:R vs dist inf mem} and \ref{fig: R vs dist fin mem}
show the secret key generation rate, at the optimal
value of intensity, versus the total distance, $L = 2 L_s + L_{\rm rep}$, between Alice and Bob. In both figures, we assume $L_s$ is a fixed short distance resembling the length of the access network. We vary $L_{\rm rep}$ then to effectively increase the link distance. Figure~\ref{fig:R vs dist inf mem} shows the secret key generation rate per transmitted pulse in the source-limited regime, whereas Fig.~\ref{fig: R vs dist fin mem} represents the key rate per logical memory used in the repeater-limited regime. In both cases we consider SPSs at $p=10^{-4}$ as well as coherent decoy states. The difference in the performance of the systems relying on these sources, as expected, is low, and that again confirms the possibility, and practicality, of using the decoy-state technique for end-user devices. The cut-off security distance, i.e., the distance beyond which secure key exchange is not possible, almost doubles every time we  increase the nesting level so long as memories decoherence rates are correspondingly low. This distance at $n=0$ is about 800~km, similar to the no-memory case for the parameter values used and at $n=1$ and $n=2$, respectively, reaches around 1500~km and 2500~km. Security distances are slightly higher for the SPSs than coherent-state sources.

The slope of the curves in Fig.~\ref{fig:R vs dist inf mem} is different than that of Fig.~\ref{fig: R vs dist fin mem}. In Fig.~\ref{fig:R vs dist inf mem} curves are almost flat until they reach their cut-off distances. That has two reasons. First, in the source-limited regime, $R_{\rm QKD}$ is proportional to the constant $R_S$, whereas, it scales with $R_{\rm ent}$, which exponentially decays with $L_0$ \cite{LoPiparo:2013}, in the repeater-limited regime. Second, and this is common in both figures, in the absence of the decoherence, the fidelity of the entangled states generated by our probabilistic repeater effectively reaches a constant value once we increase the distance \cite{Razavi.Amirloo.10}. That means that the double-photon-driven error terms in the key rate are almost fixed until dark count becomes significant and the rate goes down. 

The implications on the achievable key rate is also different in the two figures. In Fig.~\ref{fig:R vs dist inf mem}, at a nominal distance of $L = 1000$~km and a source rate of $R_S = 1$~GHz, the key rate is in the region of Mb/s. The assumption $2R_S < R_{\rm rep}$, however, implies that we need something on the order of $10^{15}$ QMs in our core network to work in the source-limited regime, which seems, at the moment, quite impractical. In the repeater-limited regime, we still need many memories to obtain a decent rate. For instance, at $L=1000$~km, we would need around 1 billion QMs to get a key rate on the order of kb/s. This is still a huge number of resources for the current technology of QMs. This is in fact the same number of memories in use in our classical computers, which was perhaps inconceivable a few decades ago. Progress in solid-state QMs is much needed to meet the above requirements.
  
\subsection{Crossover distance}

\begin{figure}
\begin{centering}
\includegraphics[width=8.6cm]{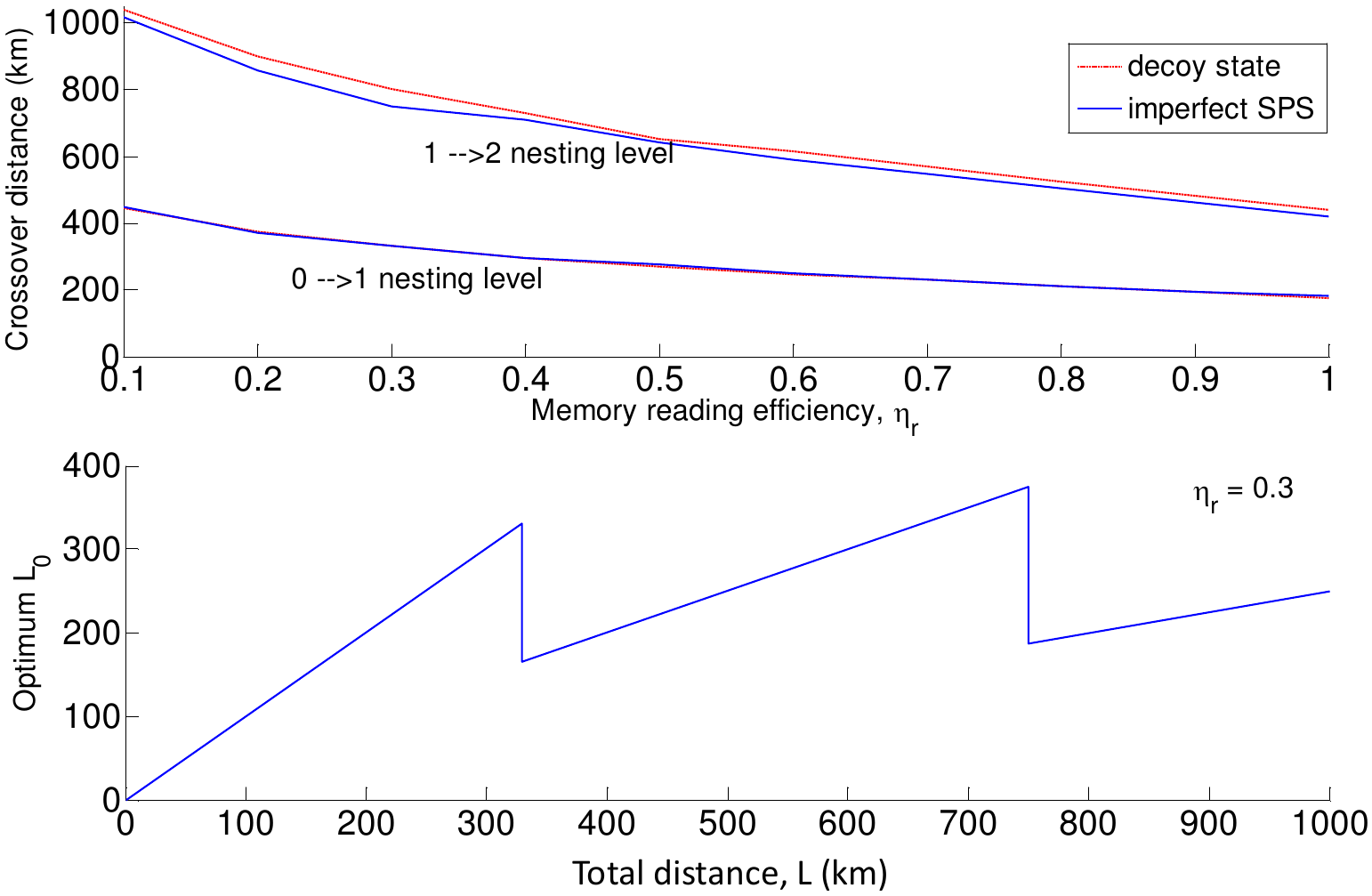}
\par\end{centering}
\caption{{\footnotesize \label{fig:(a)-Crossover-distance}(a) Crossover distance
versus QM's recall efficiency in the repeater-limited regime. 
(b) Optimum spacing $L_{0}$ between adjacent nodes of a quantum repeater at $\eta_r = 0.3$. }}
\end{figure}

The different slopes in Figs.~\ref{fig:R vs dist inf mem} and \ref{fig: R vs dist fin mem} result in appreciably different values for crossover distances, i.e., the distances where one nesting level outperforms its previous one. In the source-limited regime, in Fig.~\ref{fig:R vs dist inf mem}, the curve for $n=1$ outperforms that of $n=0$ for $L$ greater than around 750~km. The crossover distance to nesting level 2 is then around 1400~km. These are quite large distances, which imply that $L_0$, the spacing between adjacent nodes in our quantum repeater, could be as large as 700~km. This sparse location of memories in the system has some advantages in the sense that resources are more or less centralized, rather than distributed, but at the same time it imposes harder conditions on maintaining phase and polarization stability over such long distances. In the repeater-limited regime of Fig.~\ref{fig: R vs dist fin mem}, the nodes are much closer as now the crossover distance is around/below 500~km. This implies that the optimum architecture of our core network relies on, among other things, how many QMs are available at the time of development. 

The crossover distance is also a function of the efficiency of various system parameters. In Fig.~\ref{fig:(a)-Crossover-distance}(a), we have looked at the crossover distance as a function of the recall efficiency, $\eta_r$, in the repeater-limited regime. This is particularly important, because $\eta_r$ implicitly accounts for the amplitude decay in memories. As expected, the crossover distance decreases with the recall efficiency as there would be less of rate reduction because of the BSM operation. Figure~\ref{fig:(a)-Crossover-distance}(b) shows this effect on the optimal value of $L_0$. It can be seen that at $\eta_r = 0.3$ the optimal spacing is much wider than what can be obtained from Fig.~\ref{fig: R vs dist fin mem} at $\eta_r = 0.87$. It can be seen that the curve for optimal $L_0$ is non-continuous as we have limited our study to the case when the number of segments in a repeater setup is a power of 2. By developing new repeater protocols for arbitrarily number of segments, one can get a smoother curve for optimal $L_0$. At $\eta_r = 0.3$, $L_0$ is on average around 250~km for the set of parameters as in Table~\ref{tab:Nominal-values-used}.

\section{Conclusions}

In this paper we combined MDI-QKD with a quantum repeater setup in order to obtain a long-distance key exchange scheme without the need to trust any of the intermediate nodes or measurement tools. This trust-free network could be used in future generations of quantum networks, where the easy cost-efficient access to the network would be facilitated by laser-based encoders and the repeater technology, at the backbone, would be maintained by the service provider. We considered a particular entanglement distribution scheme for our quantum repeater, which relied on imperfect single-photon sources. We merged memories entangled by this probabilistic repeater setup with photons sent and phase encoded by the two users via two BSM modules. We showed that it would be possible to exchange secret keys up to over 2500~km using repeaters with two nesting levels. It turned out that in order to get a key rate on the order of 1~kb/s, one may need to employ and control billions of memories at the core network. We also showed that the network architecture depends on the number of memories at stake. In the limit of infinitely many memories, the repeater nodes would be sparsely located, although each node may contain a large number of memories. Our results showed how challenging it would be to build trust-free quantum communication networks.


%

\appendix
\section{Derivation of key rate terms}
\label{AppA}

In this Appendix, we derive the key rate terms in Eqs.~\eqref{eq:Rss} and \eqref{eq:Rcc} under the normal mode of operation when no eavesdropper is present. We use the formulation developed in Eqs.~\eqref{eq:initial_dens_mat}-\eqref{eq:measurement_operators-1} to obtain $\Gamma_{11}^z = Y_{11}^z$, $\epsilon_{11}^x = e_{11}^x$, $\Gamma_{pp}^z = Q_{pp}^z$, $\epsilon_{pp}^z = E_{pp}^z$, $\Gamma_{\mu \nu}^z = Q_{\mu \nu}^z$, and $\epsilon_{\mu \nu}^z = E_{\mu \nu}^z$, where new unifying notations $\Gamma$ and $\epsilon$ are used in this section.

\begin{table}
\begin{centering}
\vspace{2mm}
\begin{tabular}{|c|c|}
\hline 
$\rho_{AB}$  & ${\rm tr}\left(M_{x_{0}}B_{\eta_{a}\eta_{b}}^{AB}\left(\rho_{AB}\right)\right)$\tabularnewline
\hline 
\hline 
$|\alpha0\left\rangle \left\langle \alpha0|\right.\right.$ & $(1-d_{c})\left[e^{-\frac{\eta_{a}}{2}\mu}\left(1-e^{-\frac{\eta_{a}}{2}\mu}\right)+d_{c}e^{-\eta_{a}\mu}\right]$\tabularnewline
\hline 
$|\alpha1\left\rangle \left\langle \alpha1|\right.\right.$ & $(1-d_{c})\left[\frac{\eta_{b}}{2}e^{-\frac{\eta_{a}}{2}\mu}\left(1+\frac{\eta_{a}}{2}\mu\right)\right.$\tabularnewline
 & $+e^{-\frac{\eta_{a}}{2}\mu}\left(1-\eta_{b}\right)\left(1-e^{-\frac{\eta_{a}}{2}\mu}\right)$\tabularnewline
 & $\left.+d_{c}\left(1-\eta_{b}\right)\left(1-e^{-\eta_{a}\mu}\right)\right]$\tabularnewline
\hline 
$|\alpha2\left\rangle \left\langle \alpha2|\right.\right.$ & $(1-d_{c})\left\{ \frac{\eta_{b}^{2}}{4}e^{-\frac{\eta_{a}}{2}\mu}\left[1+\right.\right.$\tabularnewline
 & $+\frac{\eta_{a}^{2}}{4}\mu^{2}\left(\frac{1}{2}-8\, e^{-\frac{\eta_{a}}{2}\mu}\right)\left.+\eta_{a}\mu\right]$\tabularnewline
 & $+\eta_{b}e^{-\frac{\eta_{a}}{2}\mu}\left(1-\eta_{b}\right)\left(1+\frac{\eta_{a}}{2}\mu\right)$\tabularnewline
 & $+e^{-\frac{\eta_{a}}{2}\mu}\left(1-\eta_{b}\right)^{2}\left(1-e^{-\frac{\eta_{a}}{2}\mu}\right)$\tabularnewline
 & $\left.+d_{c}\left[\frac{\eta_{a}^{2}\eta_{b}^{2}}{2}e^{-\eta_{a}\mu}\mu^{2}+e^{-\eta_{a}\mu}\left(1-\eta_{b}\right)^{2}\right]\right\} $\tabularnewline
\hline 
$|\alpha1\left\rangle \left\langle \alpha0|\right.\right.$ & $(1-d_{c})\left(\frac{1}{2}\sqrt{\eta_{a}\eta_{b}}\alpha e^{-\frac{\eta_{a}}{2}\mu}\right)$\tabularnewline
\hline 
$|\alpha0\left\rangle \left\langle \alpha1|\right.\right.$ & $(1-d_{c})\left(\frac{1}{2}\sqrt{\eta_{a}\eta_{b}}\alpha e^{-\frac{\eta_{a}}{2}\mu}\right)$\tabularnewline
\hline 
$|\alpha1\left\rangle \left\langle \alpha2|\right.\right.$ & $(1-d_{c})\left(\sqrt{\frac{\eta_{a}\eta_{b}}{2}}\alpha\left(\frac{\eta_{b}}{2}-\frac{\eta_{a}\eta_{b}}{8}-1\right)\right)$\tabularnewline
\hline 
$|\alpha2\left\rangle \left\langle \alpha1|\right.\right.$ & $(1-d_{c})\left(\sqrt{\frac{\eta_{a}\eta_{b}}{2}}\alpha\left(\frac{\eta_{b}}{2}-\frac{\eta_{a}\eta_{b}}{8}-1\right)\right)$\tabularnewline
\hline 
\end{tabular}
\par\end{centering}

\caption{\label{tab:coherent}{\footnotesize The input-output relationship
for a butterfly module with coherent states in one input and number states in the other. The column on the right represents the probability that the output state causes a click on detector $x_0$, but not $x_1$, assuming that detector $x_0$ measures the left output port and $x_1$ the right one. The expression ${\rm tr}\left(M_{x_{1}}B_{\eta_a,\eta_b}^{AB}\left(\rho_{AB}\right)\right)$ will give the same results as above for symmetrical input states; a minus sign correction is needed for asymmetrical input states. Here, $\mu = |\alpha|^2$.} }
\end{table}
Let $\rho_{\rm enc}^\Phi(mn)$ denote the output state of Alice and Bob's encoders for, respectively, sending bits $m$ and $n$, for $m,n = 0,1$, in basis $\Phi$. With the above notation, the probability that an acceptable click pattern occurs in basis $\Phi$, $\Gamma_{\gamma \delta}^{\Phi}$, is given by 
\begin{equation}
\Gamma_{\gamma \delta}^{\Phi}= \sum_{i,j,k,l,m,n=0,1}{P_{r_{i}s_{j}u_{k}v_{l}}(\rho_{\rm enc}^\Phi(mn))/4}, 
\end{equation}
where $\gamma = \delta = 1$ refers to the case when Alice and Bob are sending exactly one photon each; when $\gamma =\delta =p$, imperfect SPSs are used and when $\gamma=\mu$ and $\delta=\nu$ coherent states with mean photon number $\mu$
and $\nu$, are, respectively, in use. In above, some of the successful click patterns would result in errors in the end, while the other in correct sifted key bits. By separating these two components, we obtain 
\begin{equation}
\Gamma_{\gamma \delta}^{\Phi}=\Gamma_{\gamma \delta;C}^{\Phi} + \Gamma_{\gamma \delta;E}^{\Phi}, 
\label{eq:y_z}
\end{equation}
where $\Gamma_{\gamma \delta;C(E)}^{\Phi}$ represents the click terms that result in correct (erroneous) inference of bits by Alice and Bob. In the $z$ basis,
\begin{equation}
\Gamma_{\gamma \delta;C}^{z}= \sum_{i,j,k,l, m,n=0,1; m+n =1}{P_{r_{i}s_{j}u_{k}v_{l}}(\rho_{\rm enc}^z(mn))/4}
\end{equation}
and $\Gamma_{\gamma \delta;E}^{\Phi} = \Gamma_{\gamma \delta}^{\Phi} - \Gamma_{\gamma \delta;C}^{\Phi}$. In the $x$ basis,
\begin{equation}
\begin{array}{c}
\Gamma_{\gamma\delta;C}^{x}={\displaystyle \sum_{i,k,m,n=0,1;m\oplus n=0}\left(P_{r_{i}s_{i}u_{k}v_{k}}(\rho_{{\rm enc}}^{x}(mn))/4\right.}\\
\left.+P_{r_{i}s_{i\oplus1}u_{k}v_{k\oplus1}}(\rho_{{\rm enc}}^{x}(mn))/4\right)\\
+{\displaystyle \sum_{i,k,m,n=0,1;m\oplus n=1}}\left(P_{r_{i}s_{i}u_{k}v_{k\oplus1}}(\rho_{{\rm enc}}^{x}(mn))/4\right.\\
\left.+P_{r_{i}s_{i\oplus1}u_{k}v_{k}}(\rho_{{\rm enc}}^{x}(mn))/4\right),
\end{array}
\end{equation}
where $\oplus$ denotes addition modulo two. Finally, all QBER terms can be obtained from the following.
\begin{equation}
\epsilon_{\gamma \delta}^{\Phi}=\frac{\Gamma_{\gamma \delta;E}^{\Phi}}{\Gamma_{\gamma \delta}^{\Phi}}.
\end{equation}
\bibliographystyle{apsrev4-1}
\bibliography{Bibli28Sept12,bib}

\end{document}